\documentclass[twocolumn,showpacs,pra,aps,eqsecnum]{revtex4}

\usepackage{amsmath}
\usepackage{amssymb}
\usepackage{graphicx}
\usepackage{upgreek}
\usepackage{bm}

\renewcommand{\vec}[1]{\bm{#1}}

\newcommand{\dif}{\mathrm{d}}
\newcommand{\mi}{\mathrm{i}}
\newcommand{\me}{\mathrm{e}}

\newcommand{\sqe}{\sqrt{\varepsilon}}
\newcommand{\ep}{\varepsilon}

\begin{document}

\title{Casimir--Polder interaction of fullerene molecules with
surfaces}

\author{Stefan Yoshi Buhmann}
\author{Stefan Scheel}
\affiliation{Quantum Optics and Laser Science, Blackett Laboratory,
Imperial College London, Prince Consort Road, London SW7 2BW, United
Kingdom}
\author{Simen {\AA}. Ellingsen}
\affiliation{Department of Energy and Process Engineering, Norwegian
University of Science and Technology, N-7491 Trondheim, Norway}
\author{Klaus Hornberger}
\author{Andreas Jacob}
\affiliation{Unversit\"at Duisburg--Essen, Fakult\"at f\"ur Physik,
Lotharstra\ss e 1-21, 47057 Duisburg, Germany}

\date{\today}

\begin{abstract}
We calculate the thermal Casimir--Polder potential of
$\mathrm{C}_{60}$ and $\mathrm{C}_{70}$ fullerene molecules near gold
and silicon nitride surfaces, motivated by their relevance for
molecular matter wave interference experiments. We obtain the
coefficients governing the asymptotic power laws of the interaction in
the thermal, retarded and nonretarded distance regimes and evaluate
the full potential numerically. The interaction is found to be
dominated by electronic transitions, and hence independent of the
internal temperature of the molecules. The contributions from phonon
transitions, which are affected by the molecular temperature, give
rise to only a small correction. Moreover, we find that the sizeable
molecular line widths of thermal fullerenes may modify the nonretarded
interaction, depending on the model used. Detailed measurements of the
nonretarded potential of fullerene thus allow one to distinguish
between different theories of incorporating damping.
\end{abstract}

\maketitle


\section{Introduction}
\label{Sec1}

It is a remarkable feature of fullerene buckyballs that they can exist
as delocalized quantum waves \cite{Arndt1999a}, as is proven almost
routinely in matter wave interference experiments
\cite{Hornberger2012a}. Such interferometer setups involve
nanomechanical grating structures, typically made of gold or silicon
nitride. As the molecules pass through the grating slits they
experience an attractive dispersion force between the polarisable
molecule and the grating wall. Even though this Casimir--Polder  (CP)
interaction is weak, it must be accounted for in predictions of the
interference fringes \cite{Grisenti1999,Nimmrichter2008a}.

The influence of dispersion forces is particularly strong in modern
near-field interference setups where many different interference
orders contribute resonantly, implying that even tiny distortions of
the molecular wave fronts affect the fringe pattern
\cite{Nimmrichter2008a}. In these experiments, it is the presence of
the  Casimir--Polder interaction which impedes the demonstration of
interference with even larger and more polarisable particles
\cite{Gerlich2007}. At the same time, this strong sensitivity of the
fullerene matter waves provides a means of verifying the precise value
and functional form of the dispersion forces. It is therefore
important to have a reliable description of the expected
Casimir--Polder potential available, which should also account for
molecules not in thermal equilibrium with their environment, given
that the beam is usually produced by thermal sublimation.

Casimir--Polder interactions have been studied intensively in recent
years, though mainly focused on atoms \cite{Scheel08}. Such studies
include thermal equilibrium \cite{Henkel02} as well as non-equilibrium
situations \cite{ducloy06,buhmann08} in which the internal temperature
of the microscopic (atomic or molecular) system can be vastly
different from that of the macroscopic environment. Carbon-based
nanostructures have been of particular interest due to applications.
In this context, the interaction of a carbon nanotube with a surface
has been studied \cite{Bordag06}.

Motivated by the mentioned matter-wave interference experiments
we now discuss and evaluate in detail the CP interaction of
$\mathrm{C}_{60}$ and $\mathrm{C}_{70}$ fullerenes with planar
surfaces made of gold or silicon nitride ($\mathrm{SiN}_x$ or
$\mathrm{Si}_3\mathrm{N}_4$). In Sect.~\ref{Sec2}, we summarise the
basic equations that govern the CP potential, determine the molecular
polarisabilities of the fullerenes from spectroscopic data, and list
the material parameters of the surface materials. In Sect.~\ref{Sec3},
we calculate the CP potentials by both numerical and analytical means
and discuss our results. A short summary is given in Sect.~\ref{Sec4}.


\section{Basic equations and parameters}
\label{Sec2}

We begin by presenting the theory of the thermal CP potential and by
recording the molecular and material properties as obtained from
optical data.


\subsection{Thermal Casimir--Polder potential}
\label{Sec2A}

We consider a non-magnetic, isotropic molecule of internal
temperature $T_\mathrm{m}$ placed at a distance $z$ from a plane
non-magnetic surface of permittivity $\varepsilon(\omega)$, with both
surface and environment being held at uniform temperature $T$. As
shown in Refs.~\cite{buhmann08} and \cite{ellingsen09}, the thermal CP
potential of the molecule can be given as a sum of non-resonant and
resonant contributions,
\begin{equation}
\label{eq1}
U(z)=U_\mathrm{nres}(z)+U_\mathrm{res}(z).
\end{equation}

The non-resonant contribution is due to virtual photons and it is
given by a sum
\begin{multline}
\label{eq2}
U_\mathrm{nres}(z)=\frac{\mu_0k_\mathrm{B}T}{8\pi}
 \sideset{}{'}\sum_{j=0}^\infty
[\alpha_{T_{\mathrm m}}(\mi\xi_j)
 +\alpha_{T_{\mathrm m}}(-\mi\xi_j)]\\
\times\int_{\xi_j/c}^\infty\dif\kappa^\perp\,
 \me^{-2\kappa^\perp z}
 \bigl[\xi_j^2r_s(\xi_j,\kappa^\perp)\\
 -\bigl(2\,\kappa^{\perp 2}c^2-\xi_j^2\bigr)
 r_p(\xi_j,\kappa^\perp)\bigr]
\end{multline}
over the purely imaginary Matsubara frequencies $\mi\xi_j$ with
$\xi_j=(2\pi k_\mathrm{B}T/\hbar)j$, where the prime indicates that
the $j=0$ term carries half-weight. The properties of the molecule are
represented by its thermal polarisability
\begin{equation}
\label{eq3}
\alpha_{T_{\mathrm m}}(\omega)
=\sum_np_n(T_{\mathrm m})\alpha_n(\omega)\;,
\end{equation}
where
\begin{equation}
\label{eq4}
p_n(T_{\mathrm m})=\frac{\me^{-E_n/(k_\mathrm{B}T_{\mathrm m})}}
 {\sum_k\me^{-E_k/(k_\mathrm{B}T_{\mathrm m})}}
\end{equation}
denotes the populations of the molecular eigenstates with energies
$E_n$ and
\begin{equation}
\label{eq5}
\alpha_n(\omega)=
 \lim_{\epsilon\to 0_+}\frac{2}{3\hbar}\sum_k
 \frac{\omega_{kn}|\vec{d}_{nk}|^2}
 {\omega_{kn}^2-\omega^2
 -\mi\omega(\Gamma_n+\Gamma_k)/2}
\end{equation}
[$\omega_{kn}\!=\!(E_k\!-\!E_n)/\hbar$: molecular transition
frequencies, $\vec{d}_{nk}$: electric dipole matrix elements,
$\Gamma_n$: level widths/damping constants, $\Gamma_0\!=\!0$]
are the associated polarisabilities. The material properties of the
surface enter via the reflection coefficients for $s$- and
$p$-polarised waves
\begin{gather}
\label{eq6}
r_s(\xi,\kappa^\perp)
=\frac{\kappa^\perp-\kappa^\perp_1}{\kappa^\perp+\kappa^\perp_1}\,,
\qquad
r_p(\xi,\kappa^\perp)
=\frac{\varepsilon(\mi\xi)\kappa^\perp-\kappa^\perp_1}
 {\varepsilon(\mi\xi)\kappa^\perp+\kappa^\perp_1}
\end{gather}
with $\kappa^\perp_1=\sqrt{\kappa^{\perp 2}%
\!+\![\varepsilon(\mi\xi)\!-\!1]\xi^2/c^2}$.

The resonant contribution to the potential is due to the absorption
and stimulated emission of real photons, it reads
\begin{multline}
\label{eq7}
U_\mathrm{res}(z)\\
=\frac{\mu_0}{12\pi}
\sum_np_n(T_{\mathrm m})
 \Biggl\{\sum_{k<n}[n_T(\omega_{nk})+1]
 -\sum_{k>n}n_T(\omega_{kn})\Biggr\}\\
\times\omega_{nk}^2|\vec{d}_{nk}|^2
 \int_0^\infty\dif k^\parallel\,\frac{k^\parallel}{k^\perp}
\biggl\{\operatorname{Im}\bigl[\me^{2\mi k^\perp z}
 r_s(|\omega_{kn}|,k^\parallel)\bigr]\\
 -\biggl(2\,\frac{k^{\perp 2}c^2}{\omega_{nk}^2}-1\biggr)
 \operatorname{Im}\bigl(\me^{2\mi k^\perp z}
 r_p(|\omega_{kn}|,k^\parallel)\bigr]\biggr\}
\end{multline}
with 
\begin{equation}
\label{eq8}
n_T(\omega)=\frac{1}{\me^{\hbar\omega/(k_\mathrm{B}T)}-1}
\end{equation}
denoting the thermal photon number. The reflection coefficients for
real frequencies can be given as
\begin{gather}
\label{eq9}
r_s(\omega,k^\parallel)
=\frac{k^\perp-k^\perp_1}{k^\perp+k^\perp_1}\,,
\qquad
r_p(\omega,k^\parallel)
=\frac{\varepsilon(\omega)k^\perp-k^\perp_1}
 {\varepsilon(\omega)k^\perp+k^\perp_1}
\end{gather}
with $k^\perp\!=\!\sqrt{\omega^2/c^2\!-\!k^{\parallel 2}}\,$ if
$k^\parallel\!\le\!\omega/c$,
$k^\perp\!=\!\mi\sqrt{k^{\parallel 2}\!-\!\omega^2/c^2}\,$ if
$k^\parallel\!\ge\!\omega/c$ and 
$k^\perp_1\!=\!\sqrt{\varepsilon(\omega)\omega^2/c^2%
\!-\!k^{\parallel 2}}\,$ with
$\operatorname{Im}\,k^\perp_1\!>\!0$.

We stress that the molecule is out of thermal equilibrium with its
environment whenever $T_\mathrm{m}\neq T$. Here, the thermal
polarisability as well as the internal-state populations of the
molecule depend on its internal temperature $T_\mathrm{m}$, while the
Matsubara frequencies and thermal photon numbers are given in terms of
the environment temperature $T$.


\subsection{Molecular and material properties}
\label{Sec2B}

The dielectric permittivity $\varepsilon(\omega)$ of thin fullerene
films in the optical regime has been measured by means of electron
energy-loss spectroscopy \cite{sohmen92} as well as in the gas phase
\cite{Ju93}. The frequencies and dipole
matrix elements of the corresponding electronic transitions of single
fullerene molecules can be deduced from the data in
Ref.~\cite{sohmen92} in a two-step procedure.

\begin{figure}[!t!]
\noindent\vspace*{-2ex}
\begin{center}
\includegraphics[width=\linewidth]{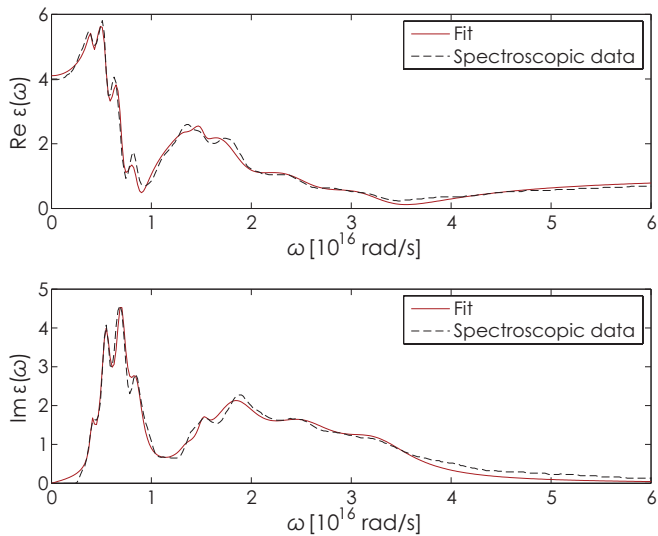}
\end{center}
\caption{
\label{Fig1}
$9$-oscillator fit (solid line) to the spectroscopic data (dashed
line) of the permittivity of a thin $\mathrm{C}_{60}$ film.
}
\end{figure}

\begin{figure}[!t!]
\noindent\vspace*{-2ex}
\begin{center}
\includegraphics[width=\linewidth]{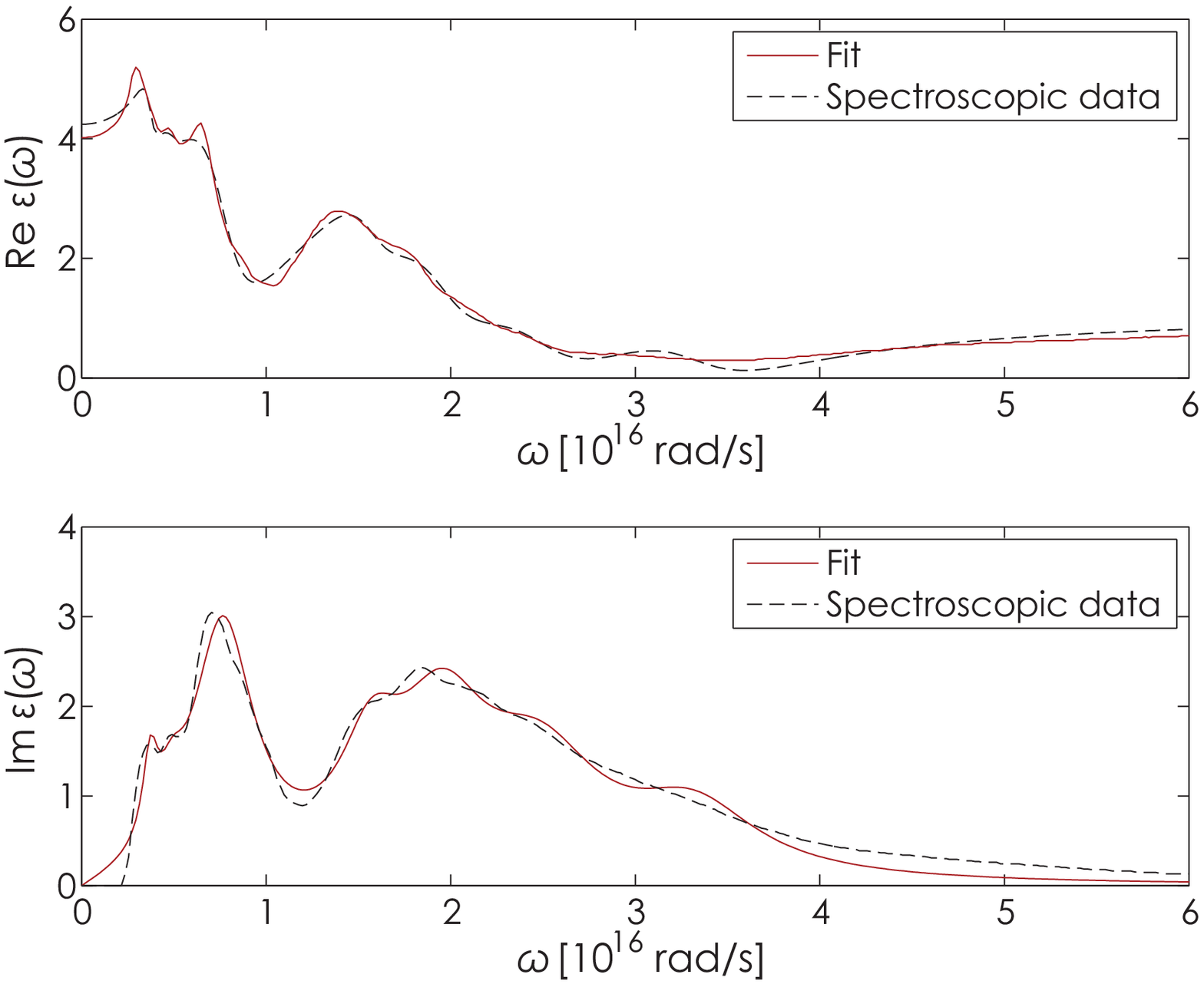}
\end{center}
\caption{
\label{Fig2}
$7$-oscillator fit (solid line) to the spectroscopic data (dashed
line) of the permittivity of a thin $\mathrm{C}_{70}$ film.
}
\end{figure}

We first apply a simultaneous least-squares fit of an $n$-oscillator
model
\begin{equation}
\label{eq10}
\varepsilon(\omega)=\varepsilon_\infty + \sum_{i=1}^n\frac{f_i\Omega_i^2}
 {\Omega_i^2-\omega^2-\mi\gamma_i\omega}
\end{equation}
to the measured data for the real and imaginary parts of the
permittivity as taken from Ref.~\cite{sohmen92}. The fits with $n=9$
oscillators for $\mathrm{C}_{60}$ and $n=7$ for $\mathrm{C}_{70}$ are
illustrated in Figs.~\ref{Fig1} and \ref{Fig2}.
The obtained fit parameters $\Omega_i$, $f_i$ and $\Gamma_i$ are given
in Tabs.~\ref{Tab1} and \ref{Tab2}.
\begin{table}[t]
\begin{center}
\begin{tabular}{|c|c|c|}
\hline
\quad\,$\Omega_i$ [$\mathrm{rad}/\mathrm{s}$]\quad\,
&\quad\,$f_i$\quad\,
&\quad\,$\gamma_i$ [$\mathrm{rad}/\mathrm{s}$]\quad\,\\
\hline
\quad\,$4.10\times 10^{15}$\quad\,
 &\quad\,$0.120$\quad\,
 &\quad\,$5.44\times 10^{14}$\quad\,\\
\quad\,$5.47\times 10^{15}$\quad\,
 &\quad\,$0.663$\quad\,
 &\quad\,$1.14\times 10^{15}$\quad\,\\
\quad\,$6.99\times 10^{15}$\quad\,
 &\quad\,$0.664$\quad\,
 &\quad\,$1.28\times 10^{15}$\quad\,\\
\quad\,$8.51\times 10^{15}$\quad\,
 &\quad\,$0.348$\quad\,
 &\quad\,$1.60\times 10^{15}$\quad\,\\
\quad\,$1.35\times 10^{16}$\quad\,
 &\quad\,$0.0270$\quad\,
 &\quad\,$1.72\times 10^{15}$\quad\,\\
\quad\,$1.52\times 10^{16}$\quad\,
 &\quad\,$0.0471$\quad\,
 &\quad\,$1.33\times 10^{15}$\quad\,\\
\quad\,$1.85\times 10^{16}$\quad\,
 &\quad\,$0.554$\quad\,
 &\quad\,$6.29\times 10^{15}$\quad\,\\
\quad\,$2.54\times 10^{16}$\quad\,
 &\quad\,$0.403$\quad\,
 &\quad\,$9.28\times 10^{15}$\quad\,\\
\quad\,$3.27\times 10^{16}$\quad\,
 &\quad\,$0.229$\quad\,
 &\quad\,$9.31\times 10^{15}$\quad\,\\
\hline
\end{tabular}
\end{center}
\caption{
\label{Tab1}
Fit parameters obtained for the permittivity of $\mathrm{C}_{60}$
films in the optical regime. $\varepsilon_\infty=1.0463$}
\end{table}
\begin{table}[t]
\begin{center}
\begin{tabular}{|c|c|c|}
\hline
\quad\,$\Omega_i$ [$\mathrm{rad}/\mathrm{s}$]\quad\,
&\quad\,$f_i$\quad\,
&\quad\,$\gamma_i$ [$\mathrm{rad}/\mathrm{s}$]\quad\,\\
\hline
\quad\,$3.76\times 10^{15}$\quad\,
 &\quad\,$0.245$\quad\,
 &\quad\,$9.91\times 10^{14}$\quad\,\\
\quad\,$4.97\times 10^{15}$\quad\,
 &\quad\,$0.170$\quad\,
 &\quad\,$1.70\times 10^{15}$\quad\,\\
\quad\,$7.92\times 10^{15}$\quad\,
 &\quad\,$1.39$\quad\,
 &\quad\,$4.05\times 10^{15}$\quad\,\\
\quad\,$1.58\times 10^{16}$\quad\,
 &\quad\,$0.276$\quad\,
 &\quad\,$4.05\times 10^{15}$\quad\,\\
\quad\,$1.98\times 10^{16}$\quad\,
 &\quad\,$0.502$\quad\,
 &\quad\,$6.17\times 10^{15}$\quad\,\\
\quad\,$2.51\times 10^{16}$\quad\,
 &\quad\,$0.393$\quad\,
 &\quad\,$8.19\times 10^{15}$\quad\,\\
\quad\,$3.34\times 10^{16}$\quad\,
 &\quad\,$0.186$\quad\,
 &\quad\,$8.45\times 10^{15}$\quad\,\\
\hline
\end{tabular}
\end{center}
\caption{
\label{Tab2}
Fit parameters obtained for the permittivity of $\mathrm{C}_{70}$
films in the optical regime. $\varepsilon_\infty=1.0827$}
\end{table}

Next, we relate the fitted permittivity to the molecular
polarisability by means of the Clausius--Mosotti law
\begin{equation}
\label{eq11}
\alpha(\omega)=\frac{3\varepsilon_0}{\eta}\,
 \frac{\varepsilon(\omega)-1}{\varepsilon(\omega)+2}
\end{equation}
where $\eta$ is the number density of fullerene molecules in the
film. For a face-centred-cubic crystal structure of the fullerene
molecules in the thin film, one has $\eta=4/a^3$ with a
lattice constants $a\!=\!1.42\times 10^{-9}\mathrm{m}$ for
$\mathrm{C}_{60}$ and $a\!=\!1.51\times 10^{-9}\mathrm{m}$ for
$\mathrm{C}_{70}$ \cite{sohmen92}. We employ a decomposition of the
resulting expression into partial fractions to write it in the form
\begin{equation}
\label{eq12}
\alpha_0(\omega)=
 \frac{2}{3\hbar}\sum_k
 \frac{\omega_{k0}|\vec{d}_{0k}|^2}
 {\omega_{k0}^2-\omega^2-\mi\omega\Gamma_k/2}\,,
\end{equation}
from which the transition frequencies $\omega_{k0}$, dipole matrix
elements $d_{0k}$ and excited-state widths $\Gamma_k$ can be read off.
They are given in Tabs.~\ref{Tab3} and \ref{Tab4}. Note that we have
used the ground-state polarisability $\alpha_0(\omega)$ rather than
its thermal counterpart~(\ref{eq3}). This is a good approximation for
the considered electronic transitions whose frequencies are much
higher than the thermal frequency
$k_\mathrm{B}T_\mathrm{m}/\hbar\!=\!3.93\times%
10^{13}\,\mathrm{rad}/\mathrm{s}$ corresponding to the temperature of
the molecules in the experiment \cite{sohmen92},
$T_\mathrm{m}=300\,\mathrm{K}$.
\begin{table}[t]
\begin{center}
\begin{tabular}{|c|c|c|}
\hline
\quad\,$\omega_{k0}$ [$\mathrm{rad}/\mathrm{s}$]\quad\,
&\quad\,$d_{0k}$ [$\mathrm{Cm}$]\quad\,
&\quad\,$\Gamma_k$ [$\mathrm{rad}/\mathrm{s}$]\quad\,\\
\hline
\quad\,$4.14\times 10^{15}$\quad\,
 &\quad\,$7.93\times 10^{-30}$\quad\,
 &\quad\,$1.10\times 10^{15}$\quad\,\\
\quad\,$5.73\times 10^{15}$\quad\,
 &\quad\,$2.36\times 10^{-29}$\quad\,
 &\quad\,$2.32\times 10^{15}$\quad\,\\
\quad\,$7.43\times 10^{15}$\quad\,
 &\quad\,$3.58\times 10^{-29}$\quad\,
 &\quad\,$2.65\times 10^{15}$\quad\,\\
\quad\,$8.97\times 10^{15}$\quad\,
 &\quad\,$4.93\times 10^{-29}$\quad\,
 &\quad\,$3.23\times 10^{15}$\quad\,\\
\quad\,$1.36\times 10^{16}$\quad\,
 &\quad\,$1.09\times 10^{-29}$\quad\,
 &\quad\,$3.47\times 10^{15}$\quad\,\\
\quad\,$1.53\times 10^{16}$\quad\,
 &\quad\,$1.65\times 10^{-29}$\quad\,
 &\quad\,$2.73\times 10^{15}$\quad\,\\
\quad\,$1.98\times 10^{16}$\quad\,
 &\quad\,$7.28\times 10^{-29}$\quad\,
 &\quad\,$1.28\times 10^{16}$\quad\,\\
\quad\,$2.70\times 10^{16}$\quad\,
 &\quad\,$9.42\times 10^{-29}$\quad\,
 &\quad\,$1.83\times 10^{16}$\quad\,\\
\quad\,$3.43\times 10^{16}$\quad\,
 &\quad\,$1.11\times 10^{-28}$\quad\,
 &\quad\,$1.85\times 10^{16}$\quad\,\\
\hline
\end{tabular}
\end{center}
\caption{
\label{Tab3}
Transition frequencies, dipole matrix elements and widths of the
electronic transitions of $\mathrm{C}_{60}$.}
\end{table}
\begin{table}[t]
\begin{center}
\begin{tabular}{|c|c|c|}
\hline
\quad\,$\omega_{k0}$ [$\mathrm{rad}/\mathrm{s}$]\quad\,
&\quad\,$d_{0k}$ [$\mathrm{Cm}$]\quad\,
&\quad\,$\Gamma_k$ [$\mathrm{rad}/\mathrm{s}$]\quad\,\\
\hline
\quad\,$3.83\times 10^{15}$\quad\,
 &\quad\,$1.40\times 10^{-29}$\quad\,
 &\quad\,$2.01\times 10^{15}$\quad\,\\
\quad\,$5.03\times 10^{15}$\quad\,
 &\quad\,$1.50\times 10^{-29}$\quad\,
 &\quad\,$3.43\times 10^{15}$\quad\,\\
\quad\,$9.04\times 10^{15}$\quad\,
 &\quad\,$6.91\times 10^{-29}$\quad\,
 &\quad\,$8.12\times 10^{15}$\quad\,\\
\quad\,$1.63\times 10^{16}$\quad\,
 &\quad\,$4.03\times 10^{-29}$\quad\,
 &\quad\,$8.28\times 10^{15}$\quad\,\\
\quad\,$2.10\times 10^{16}$\quad\,
 &\quad\,$7.90\times 10^{-29}$\quad\,
 &\quad\,$1.25\times 10^{16}$\quad\,\\
\quad\,$2.69\times 10^{16}$\quad\,
 &\quad\,$1.15\times 10^{-28}$\quad\,
 &\quad\,$1.60\times 10^{16}$\quad\,\\
\quad\,$3.48\times 10^{16}$\quad\,
 &\quad\,$1.13\times 10^{-28}$\quad\,
 &\quad\,$1.68\times 10^{16}$\quad\,\\
\hline
\end{tabular}
\end{center}
\caption{
\label{Tab4}
Transition frequencies, dipole matrix elements and widths of the
electronic transitions of $\mathrm{C}_{70}$.}
\end{table}

In addition to the optical transitions, four phonon transitions have
been identified for $\mathrm{C}_{60}$ in the infrared frequency
regime. The respective permittivity data obtained from Fourier
transform infrared experiments have been fitted to a model of the
kind given by Eq.~(\ref{eq10}) \cite{Eklund95}, with the parameters
being listed in Tab.~\ref{Tab5}.
\begin{table}[t]
\begin{center}
\begin{tabular}{|c|c|c|}
\hline
\quad\,$\Omega_i$ [$\mathrm{rad}/\mathrm{s}$]\quad\,
&\quad\,$f_i$\quad\,
&\quad\,$\gamma_i$ [$\mathrm{rad}/\mathrm{s}$]\quad\,\\
\hline
\quad\,$9.91\times 10^{13}$\quad\,
 &\quad\,$0.024$\quad\,
 &\quad\,$4.33\times 10^{11}$\quad\,\\
\quad\,$1.08\times 10^{14}$\quad\,
 &\quad\,$0.007$\quad\,
 &\quad\,$6.03\times 10^{11}$\quad\,\\
\quad\,$2.23\times 10^{14}$\quad\,
 &\quad\,$0.0011$\quad\,
 &\quad\,$5.46\times 10^{11}$\quad\,\\
\quad\,$2.69\times 10^{14}$\quad\,
 &\quad\,$0.001$\quad\,
 &\quad\,$6.40\times 10^{11}$\quad\,\\
\hline
\end{tabular}
\end{center}
\caption{
\label{Tab5}
Fit parameters for the permittivity of $\mathrm{C}_{60}$ films
in the infrared regime.}
\end{table}
To obtain the respective molecular polarisability, we make use of the
Clausius--Mosotti law~(\ref{eq11}). As the measurements have been
performed at room temperature ($T_{\mathrm m}=300\,\mathrm{K}$) where
the phonons are excited to a considerable degree, we have to employ
the thermal polarisability (\ref{eq3}). According to Eqs.~(\ref{eq4})
and (\ref{eq5}), the latter can be written in the form
\begin{multline}
\label{eq13}
\alpha_{T_{\mathrm m}}(\omega)
=\frac{2}{3\hbar}\sum_{n,k}p_n(T_{\mathrm m})\,
 \frac{\omega_{kn}|\vec{d}_{nk}|^2}
 {\omega_{kn}^2-\omega^2-\mi\omega(\Gamma_n+\Gamma_k)/2}\\
=\frac{2}{3\hbar}\sum_{n<k}p_{nk}(T_{\mathrm m})
 \tanh\biggl(\frac{\hbar\omega_{kn}}{2k_\mathrm{B}T_{\mathrm
m}}\biggr)\\
\times\frac{\omega_{kn}|\vec{d}_{nk}|^2}
 {\omega_{kn}^2-\omega^2-\mi\omega(\Gamma_n+\Gamma_k)/2}
\end{multline}
with $p_{nk}(T)\!=\!p_n(T)\!+\!p_k(T)$. Assuming that all observed
phonon transitions are from the ground state ($n\!=\!0$), we have
\begin{multline}
\label{eq14}
\alpha_{T_{\mathrm m}}(\omega)
=\frac{2}{3\hbar}\sum_{k}p_{0k}(T_{\mathrm m})
 \tanh\biggl(\frac{\hbar\omega_{k0}}
 {2k_\mathrm{B}T_\mathrm{m}}\biggr)\\
\times\frac{\omega_{k0}|\vec{d}_{0k}|^2}
 {\omega_{k0}^2-\omega^2-\mi\omega\Gamma_k/2}\;.
\end{multline}
The transition frequencies and dipole matrix elements can then be
readily obtained by comparing with the fit for Eq.~(\ref{eq11}), with
the results being given in Tab.~\ref{Tab6}.
\begin{table}[t]
\begin{center}
\begin{tabular}{|c|c|c|c|}
\hline
\quad\,$\omega_{k0}$ [$\mathrm{rad}/\mathrm{s}$]\quad\,
&\quad\,$d_{0k}$ [$\mathrm{Cm}$]\quad\,
&\quad\,$\Gamma_k$ [$\mathrm{rad}/\mathrm{s}$]\quad\,\\
\hline
\quad\,$9.95\times 10^{13}$\quad\,
 &\quad\,$1.69\times 10^{-30}$\quad\,
 &\quad\,$8.67\times 10^{11}$\quad\,\\
\quad\,$1.09\times 10^{14}$\quad\,
 &\quad\,$1.00\times 10^{-30}$\quad\,
 &\quad\,$1.21\times 10^{12}$\quad\,\\
\quad\,$2.23\times 10^{14}$\quad\,
 &\quad\,$5.33\times 10^{-31}$\quad\,
 &\quad\,$1.09\times 10^{12}$\quad\,\\
\quad\,$2.69\times 10^{14}$\quad\,
 &\quad\,$5.58\times 10^{-31}$\quad\,
 &\quad\,$1.28\times 10^{12}$\quad\,\\
\hline
\end{tabular}
\end{center}
\caption{
\label{Tab6}
Transition frequencies, dipole matrix elements and widths of the
phonon transitions of $\mathrm{C}_{60}$.}
\end{table}

We are going to study the interaction of fullerene molecules with Au,
$\mathrm{SiN}_x$ and $\mathrm{Si}_3\mathrm{N}_4$. The permittivity of
Au can be given as
\begin{equation}
\label{eq15}
\varepsilon(\omega)=1-\frac{\Omega_0^2}{\omega(\omega+\mi\gamma_0)}
+\sum_{i=1}^6\frac{f_i\Omega_i^2}
 {\Omega_i^2-\omega^2-\mi\gamma_i\omega}
\end{equation}
where the first term is the response of the conduction electrons as
described by a Drude model \cite{Lambrecht00} and the Lorentz-type
contributions are due to atomic transitions ($6$-oscillator fit
\cite{Decca07} based on data from Refs.~\cite{parsegian81} and
\cite{palik95}). Values for the model parameters as taken from the
above mentioned references are listed in Tab.~\ref{Tab7}.
\begin{table}[t]
\begin{center}
\begin{tabular}{|c|c|c|c|}
\hline
\quad\,$i$\quad\,
&\quad\,$\Omega_i$ [$\mathrm{rad}/\mathrm{s}$]\quad\,
&\quad\,$f_i$\quad\,
&\quad\,$\gamma_i$ [$\mathrm{rad}/\mathrm{s}$]\quad\,\\
\hline
\quad\,$0$\quad\,
 &\quad\,$1.37\times 10^{16}$\quad\,
 &\quad\,\quad\,
 &\quad\,$5.32\times 10^{13}$\quad\,\\
\quad\,$1$\quad\,
 &\quad\,$4.63\times 10^{15}$\quad\,
 &\quad\,$0.762$\quad\,
 &\quad\,$1.14\times 10^{15}$\quad\,\\
\quad\,$2$\quad\,
 &\quad\,$6.30\times 10^{15}$\quad\,
 &\quad\,$2.41$\quad\,
 &\quad\,$2.81\times 10^{15}$\quad\,\\
\quad\,$3$\quad\,
 &\quad\,$8.20\times 10^{15}$\quad\,
 &\quad\,$0.0926$\quad\,
 &\quad\,$1.52\times 10^{15}$\quad\,\\
\quad\,$4$\quad\,
 &\quad\,$1.29\times 10^{16}$\quad\,
 &\quad\,$2.14$\quad\,
 &\quad\,$1.06\times 10^{16}$\quad\,\\
\quad\,$5$\quad\,
 &\quad\,$2.05\times 10^{16}$\quad\,
 &\quad\,$0.244$\quad\,
 &\quad\,$9.12\times 10^{15}$\quad\,\\
\quad\,$6$\quad\,
 &\quad\,$3.27\times 10^{16}$\quad\,
 &\quad\,$0.670$\quad\,
 &\quad\,$1.37\times 10^{16}$\quad\,\\
\hline
\end{tabular}
\end{center}
\caption{
\label{Tab7}
Model parameters for the permittivity of Au.}
\end{table}

Diffraction experiments frequently use gratings made of low-pressure
chemical vapour deposited silicon nitride. The imaginary part of the
permittivity of this (non-stoichiometric) $\mathrm{SiN}_x$ has been
determined from optical measurements. As reported in
Ref.~\cite{Bruehl02}, one has
\begin{equation}
\label{eq16}
\operatorname{Im}\varepsilon(\omega)=\Theta(\omega-\Omega_T)\,
 \frac{f\Omega\gamma(\omega-\Omega_T)^2}
 {[(\omega^2-\Omega^2)^2-\gamma^2\omega^2]\omega}
\end{equation}
with parameters
$\Omega_T\!=\!3.48\times 10^{15}\,\mathrm{rad}/\mathrm{s}$,
$\Omega\!=\!1.09\times 10^{16}\,\mathrm{rad}/\mathrm{s}$,
$f\!=\!1.13\times 10^{17}\,\mathrm{rad}/\mathrm{s}$ and
$\gamma\!=\!1.16\times 10^{16}\,\mathrm{rad}/\mathrm{s}$. The
permittivity at imaginary frequencies as required for the non-resonant
CP potential can be obtained from the Kramers--Kronig relation
\begin{equation}
\label{eq17}
\varepsilon(\mi\xi)
=\frac{2}{\pi}\int_0^\infty\dif\omega\,
 \frac{\omega\operatorname{Im}\varepsilon(\omega)}
 {\omega^2+\xi^2}\;.
\end{equation}
In particular, $\varepsilon(0)=3.87$.

\begin{figure}[!t!]
\noindent\vspace*{-2ex}
\begin{center}
\includegraphics[width=\linewidth]{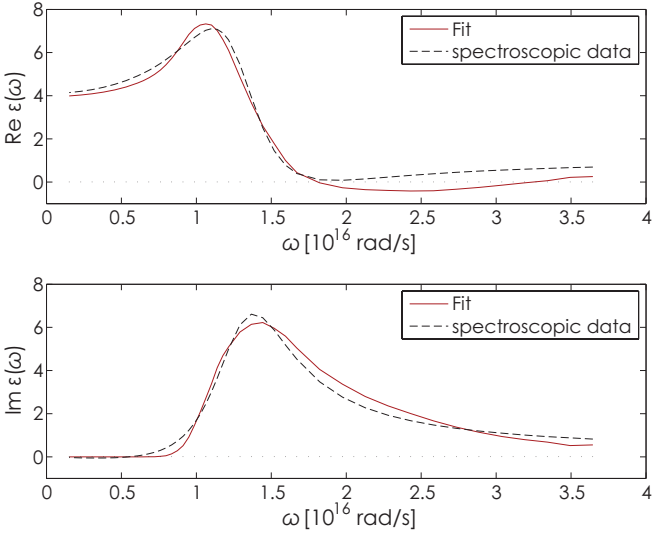}
\end{center}
\caption{
\label{Fig2b}
4-parameter semi-quantum model fit (solid line) to the spectroscopic
data (dashed line) of the permittivity $\mathrm{Si}_3\mathrm{N}_4$. 
}
\end{figure}

Alternatively, we also consider non-crystalline
$\mathrm{Si}_3\mathrm{N}_4$. The real and imaginary parts of its
permittivity have been reported over a wide frequency range
\cite{Palik91}. We have fitted this data with a single-resonance
4-parameter semi-quantum model \cite{Gervais74},
\begin{equation}
\label{eq17b}
\varepsilon(\omega)
=\frac{\Omega_\mathrm{L}^2-\omega^2-\mi\omega\gamma_\mathrm{L}}
{\Omega_\mathrm{T}^2-\omega^2-\mi\omega\gamma_\mathrm{T}}\,.
\end{equation}
The fit, as displayed in Fig.~\ref{Fig2b}, yields the parameters
$\Omega_L\!=\!2.69\times 10^{16}\,\mathrm{rad}/\mathrm{s}$,
$\Omega_T\!=\!1.33\times 10^{16}\,\mathrm{rad}/\mathrm{s}$,
$\gamma_L\!=\!3.05\times 10^{16}\,\mathrm{rad}/\mathrm{s}$, and
$\gamma_T\!=\!6.40\times 10^{15}\,\mathrm{rad}/\mathrm{s}$.
This yields a static permittivity $\varepsilon(0)=4.10$.


\section{Casimir--Polder potential of fullerene}
\label{Sec3}

Using the basic formulas from Sect.~\ref{Sec2A} together with the
molecular and material parameters from Sec.~\ref{Sec2B}, we can now
evaluate the CP potential. We begin by calculating the CP potential
associated with electronic transitions. As seen from Tabs.~\ref{Tab3}
and \ref{Tab4}, the electronic transition frequencies of
$\mathrm{C}_{60}$ and $\mathrm{C}_{70}$ are much larger than the
respective thermal frequency
$k_\mathrm{B}T/\hbar\!=\!3.93\times 10^{13}\,\mathrm{rad}/\mathrm{s}$
even at room temperature. As a consequence, the thermal photon numbers
$n_T(\omega_{nk})$ are extremely small. The molecules 
have shown interference at internal temperatures of $2500\,\mathrm{K}$
\cite{Hackermuller2004a,Hornberger05}, and they are stable up 
to $6000\,\mathrm{K}$ \cite{Mitzner95}. Even at the latter
temperature, the thermal frequency
$k_\mathrm{B}T_\mathrm{m}/\hbar\!=\!7.86\times
10^{14}\,\mathrm{rad}/\mathrm{s}$
is much smaller than the molecular transition frequencies, so the
molecule is essentially in its electronic ground state,
$p_n(T_\mathrm{m})\!=\!\delta_{n0}$. As a result, the resonant CP
potential~(\ref{eq7}) vanishes and the CP potential~(\ref{eq1}) is
entirely given by the nonresonant contribution~(\ref{eq2}) which
simplifies to
\begin{multline}
\label{eq18}
U(z)=\frac{\mu_0k_\mathrm{B}T}{8\pi}
 \sideset{}{'}\sum_{j=0}^\infty\xi_j^2
 [\alpha_0(\mi\xi_j)+\alpha_0(-\mi\xi_j)]\\
\times\int_{\xi_j/c}^\infty\dif\kappa^\perp\,
 \me^{-2\kappa^\perp z}\biggl[r_s
 -\biggl(2\,\frac{\kappa^{\perp 2}c^2}{\xi_j^2}-1\biggr)r_p\biggr]\;.
\end{multline}

As first shown in Ref.~\cite{Lifshitz56}, we may distinguish three
asymptotic regimes where the potential reduces to simple power laws.
At distances much larger than the wavelength of the predominant
thermal photons, $z\!\gg\!\hbar c/(k_\mathrm{B}T)$, the Matsubara sum
is dominated by its first term. Using Eq.~(\ref{eq6}) and carrying
out the $\kappa^\perp$-integral, we have
\begin{equation}
\label{eq19}
U(z)=-\frac{C_{3T}}{z^3}
\end{equation}
with [$\alpha_0(0)\!\equiv\!\alpha_0$]
\begin{equation}
\label{eq20}
C_{3T}
=\frac{k_\mathrm{B}T\alpha_0}{16\pi\varepsilon_0}\,
 \frac{\varepsilon(0)-1}{\varepsilon(0)+1}\;.
\end{equation}
For smaller distances $z\!\ll\!\hbar c/(k_\mathrm{B}T)$, the Matsubara
sum is well approximated by an integral, so that
\begin{multline}
\label{eq21}
U(z)=\frac{\hbar\mu_0}{16\pi^2}
\int_0^\infty\dif\xi\,\xi^2
 [\alpha_0(\mi\xi)+\alpha_0(-\mi\xi)]\\
\times\int_{\xi/c}^\infty\dif\kappa^\perp\,
 \me^{-2\kappa^\perp z}
 \biggl[r_s
 -\biggl(2\,\frac{\kappa^{\perp 2}c^2}{\xi^2}-1\biggr)r_p\biggr]\;.
\end{multline}
This distance region can be further divided into the retarded and
nonretarded regimes. At retarded distances
$c/\omega_{k0}\!\ll\!z\!\ll\!\hbar c/(k_\mathrm{B}T)$, the
approximations $\alpha_0(\mi\xi)\!\simeq\!\alpha_0$ and
$\varepsilon(\mi\xi)\!\simeq\!\varepsilon(0)\!\equiv\!\varepsilon$
lead to
\begin{equation}
\label{eq22}
U(z)=-\frac{C_4}{z^4}
\end{equation}
with
\begin{multline}
\label{eq23}
C_4=\frac{3\hbar c\alpha_0}{64\pi^2\varepsilon_0}
 \int_1^\infty\dif v\,
 \Biggl[\biggl(\frac{2}{v^2}
 -\frac{1}{v^4}\biggr)
 \frac{\varepsilon v-\sqrt{\varepsilon-1+v^2}}
 {\varepsilon v+\sqrt{\varepsilon-1+v^2}}\\
-\,\frac{1}{v^4}\,\frac{v-\sqrt{\varepsilon-1+v^2}}
 {v+\sqrt{\varepsilon-1+v^2}}\Biggr]
\end{multline}
where we have introduced the new integration variable
$v=\kappa^\perp c/\xi$. An explicit formula for $C_4$ and some of its
limits are found in App.~\ref{AppA}. The values of $\alpha_0$ are
$9.72\times 10^{-39}\mathrm{C}^2\mathrm{m}^2/\mathrm{J}$ and
$1.19\times 10^{-38}\mathrm{C}^2\mathrm{m}^2/\mathrm{J}$
for $\mathrm{C}_{60}$ and $\mathrm{C}_{70}$, respectively. For
nonretarded distances $z\!\ll\!c/\omega_{k0}$, an asymptotic expansion
in terms of $z$ leads to
\begin{equation}
\label{eq24}
U(z)=-\frac{C_3}{z^3}
\end{equation}
with
\begin{equation}
\label{eq25}
C_3=\frac{\hbar}{32\pi^2\varepsilon_0}
 \int_0^\infty\dif\xi\,
 [\alpha_0(\mi\xi)+\alpha_0(-\mi\xi)]\,
 \frac{\varepsilon(\mi\xi)-1}{\varepsilon(\mi\xi)+1}\;.
\end{equation}

Using the parameters of Sect.~\ref{Sec2B}, we have calculated the
values of the coefficients $C_3$, $C_4$ and $C_{3T}$ for
$\mathrm{C}_{60}$ and $\mathrm{C}_{70}$ molecules interacting with a
perfectly conducting surface as well as gold and silicon nitride
surfaces. The results are given in Tab.~\ref{Tab8}.
\begin{table*}[!t!]
\begin{center}
\begin{tabular}{|c||c|c|c|c|c|c|}
\hline
\quad\,Coefficient $\rightarrow$\quad\,
&\multicolumn{2}{c|}{\quad\,$C_3$ [$\mathrm{J}\mathrm{m}^3$]\quad\,}
&\multicolumn{2}{c|}{\quad\,$C_4$ [$\mathrm{J}\mathrm{m}^4$]\quad\,}
&\multicolumn{2}{c|}{\quad\,$C_{3T}$
[$\mathrm{J}\mathrm{m}^3$]\quad\,}\\
\hline
\quad\,Material $\downarrow$\quad\,
&\,$\mathrm{C}_{60}$\,
&\,$\mathrm{C}_{70}$\,
&\,$\mathrm{C}_{60}$\,
&\,$\mathrm{C}_{70}$\,
&\,$\mathrm{C}_{60}$\,
&\,$\mathrm{C}_{70}$\,\\
\hline
\hline
\quad\,Perfect conductor\quad\,
&\,$2.4\times 10^{-47}$\,
&\,$3.0\times 10^{-47}$\,
&\,$3.3\times 10^{-55}$\,
&\,$4.0\times 10^{-55}$\,
&\,$9.0\times 10^{-50}$\,
&\,$1.1\times 10^{-49}$\,\\
\quad\,Au\quad\,
&\,$1.0\times 10^{-47}$\,
&\,$1.3\times 10^{-47}$\,
&\,$3.3\times 10^{-55}$\,
&\,$4.0\times 10^{-55}$\,
&\,$9.0\times 10^{-50}$\,
&\,$1.1\times 10^{-49}$\,\\
\quad\,$\mathrm{Si}_3\mathrm{N}_4$\quad\,
&\,$8.4\times 10^{-48}$\,
&\,$1.1\times 10^{-47}$\,
&\,$1.5\times 10^{-55}$\,
&\,$1.9\times 10^{-55}$\,
&\,$5.5\times 10^{-50}$\,
&\,$6.7\times 10^{-50}$\,\\
\quad\,$\mathrm{SiN}_x$\quad\,
&\,$6.3\times 10^{-48}$\,
&\,$7.9\times 10^{-48}$\,
&\,$1.4\times 10^{-55}$\,
&\,$1.8\times 10^{-55}$\,
&\,$5.3\times 10^{-50}$\,
&\,$6.5\times 10^{-50}$\,\\
\hline
\end{tabular}
\end{center}
\caption{
\label{Tab8}
Coefficients for the asymptotic power laws of the CP potential of
fullerene ($T\!=\!300\,\mathrm{K}$).}
\end{table*}
Due to its larger dipole moments, all coefficients are larger for
$\mathrm{C}_{70}$ than they are for $\mathrm{C}_{60}$. The difference
is most pronounced in the nonretarded regime. Comparing the
coefficients for the different materials, we note that Au can be
considered a perfect metal in the retarded and thermal regimes, but
corrections due to finite reflectivity are quite significant at
nonretarded distances, leading to a reduction of $C_3$ by more than a
factor of $2$. This is due to the fact that the molecular transition
frequencies are a sizeable fraction of the Au plasma frequency. The
silicon nitride potentials are smaller than those of Au due to their
smaller permittivity. This difference is most pronounced at large
distances. Comparing the two different silicon nitride species reveals
that the coefficients for $\mathrm{Si}_3\mathrm{N}_4$ are larger than
those of $\mathrm{SiN}_x$ by up to $40\%$ in the nonretarded regime.

\begin{figure}[!t!]
\noindent\vspace*{-2ex}
\begin{center}
\includegraphics[width=\linewidth]{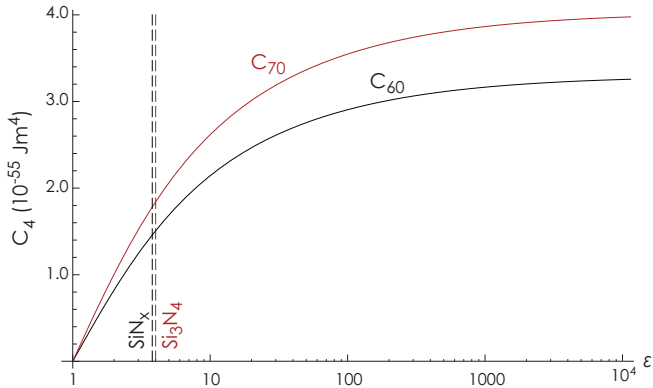}
\end{center}
\caption{
\label{Fig3}
Dependence of the retarded CP coefficient $C_4$ of $\mathrm{C}_{60}$
on the static permittivity.
}
\end{figure}

The dependence of the retarded CP coefficient $C_4$ on the static
permittivity of the surface material is displayed in Fig.~\ref{Fig3}.
The figure shows that the close similarity of the $C_4$ coefficients
for the two silicon nitride species is due to their similar static
permittivities. It also reveals that the asymptotic $C_4$ value of a
metal ($\varepsilon\!\to\!\infty$) is only reached for very large
permittivities, hence the large difference compared to Au.

The full potential of $\mathrm{C}_{60}$ in front of an Au surface has
been calculated numerically and is displayed in Fig.~\ref{Fig4}. The
tabulated results can also be found in the supplementary materials.
\begin{figure*}[!t!]
\noindent\vspace*{-2ex}
\begin{center}
\includegraphics[width=.7\linewidth]{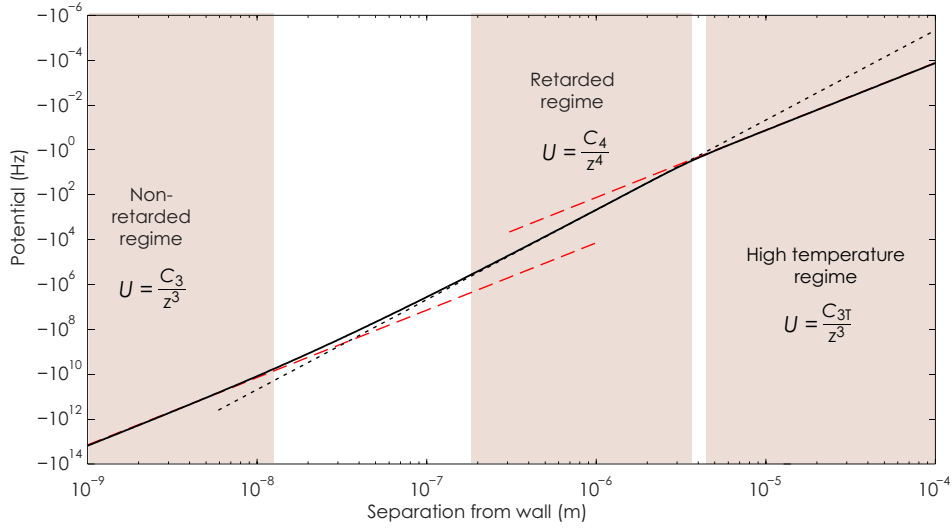}
\end{center}
\caption{
\label{Fig4}
CP potential of $\mathrm{C}_{60}$ in front of an Au surface at room
temperature and its asymptotes.
}
\end{figure*}%
The figure shows that the potential is faithfully represented by the
asymptotic power laws~(\ref{eq19}), (\ref{eq22}) and (\ref{eq24})
over a large part of the displayed distance range (as indicated by
the shaded areas). However, there is a large gap between the
nonretarded and retarded regions (between $10^{-7}\,\mathrm{m}$ and
$2\times 10^{-8}\,\mathrm{m}$) where neither limit applies.
The potentials for $\mathrm{C}_{70}$ and for different surfaces show a
similar qualitative behaviour.

\begin{figure}[!t!]
\noindent\vspace*{-2ex}
\begin{center}
\includegraphics[width=\linewidth]{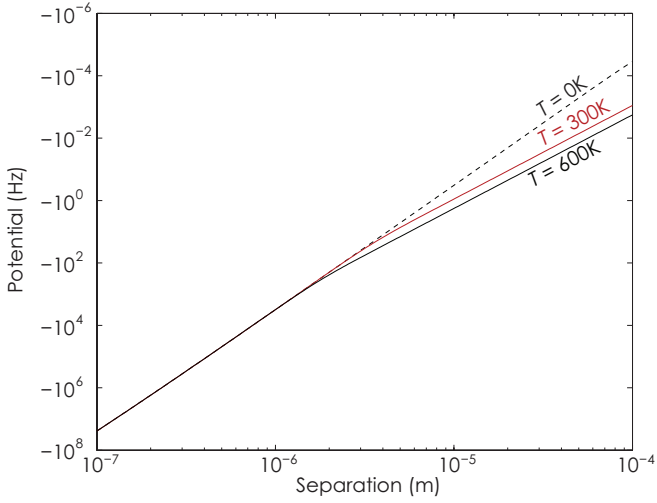}
\end{center}
\caption{
\label{Fig5}
CP potential of $\mathrm{C}_{60}$ in front of an Au surface at
different temperatures.
}
\end{figure}

The environment temperature $T$ affects the CP potential at distances
larger than the thermal wavelength where thermal photons lead to
softening of the potential decay. This temperature-dependence is
demonstrated in Fig.~\ref{Fig5} where we display the CP potential for
different environment temperatures.
As seen, thermal photons begin to affect the potential at smaller
distances for higher environment temperatures, resulting in larger
long-distance potentials. The potentials for $300\,\mathrm{K}$ and
$600\,\mathrm{K}$ begin to differ from the zero-temperature result at
distances larger than about $2\,\mu\mathrm{m}$ or $4\,\mu\mathrm{m}$,
respectively.

\begin{figure}[!t!]
\noindent\vspace*{-2ex}
\begin{center}
\includegraphics[width=\linewidth]{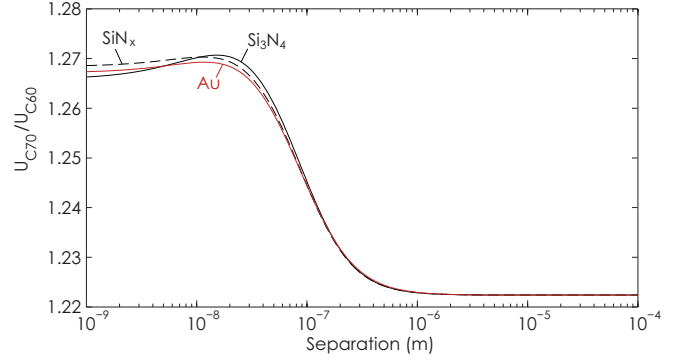}
\end{center}
\caption{
\label{Fig6}
Comparison of the CP potentials of $\mathrm{C}_{60}$ and
$\mathrm{C}_{70}$ in front of gold and silicon nitride surfaces at
room temperature.
}
\end{figure}

In Fig.~\ref{Fig6}, we compare the potentials of the two different
types of fullerenes.
The $\mathrm{C}_{70}$ potential is larger than that of
$\mathrm{C}_{60}$ by a factor which ranges between $1.3$ at small
distances and $1.2$ at large distances. This difference is due to the
larger dipole moments of $\mathrm{C}_{70}$, recall Eqs.~(\ref{eq25})
and (\ref{eq23}) for the $C_3$ and $C_4$ coefficients and (\ref{eq5})
for the atomic polarisability. The ratio of the potentials drops in
the transition region between the nonretarded and retarded regimes.
This is because the transition frequencies of $\mathrm{C}_{70}$ are
slightly larger than those of $\mathrm{C}_{60}$.

\begin{figure}[!t!]
\noindent\vspace*{-2ex}
\begin{center}
\includegraphics[width=\linewidth]{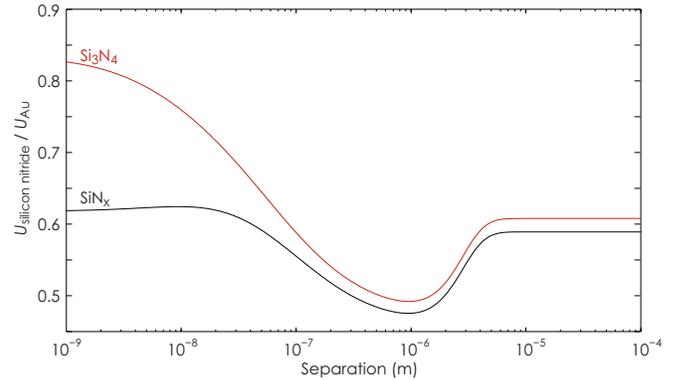}
\end{center}
\caption{
\label{Fig8}
Comparison of the CP potentials of fullerene in front of
$\mathrm{Si}_3\mathrm{N}_4$ and $\mathrm{SiN}_x$ surfaces at room
temperature, normalised with respect to the corresponding Au
potentials.
}
\end{figure}

The potentials of the two silicon nitride species are compared to that
of Au in Fig.~\ref{Fig8}. Note that the curves for the two different
fullerene molecules are indistinguishable in this plot.
We see that the ratio is roughly $60\%$ for both silicon nitride
species at large distances. This similarity is due to their very
similar static permittivities. At smaller distances, the potential of
$\mathrm{Si}_3\mathrm{N}_4$ is larger than that of
$\mathrm{SiN}_x$ by $30\%$, because its reflection coefficient falls
off less rapidly with frequency.

As seen from Tables~\ref{Tab3} and \ref{Tab4}, the line widths of the
optical transitions are almost comparable to the transition
frequencies. In our theory, the impact of the line widths is accounted
for microscopically \cite{Buhmann04}: First, one explicitly solves the
internal dynamics of the molecules, which depends on the transition
frequencies and line widths. In a second step, this solution is used
to determine the CP potential~(\ref{eq18}), where the dependence on
the molecular parameters can be expressed via the molecular
polarisability in a symmetrised form,
$\frac{1}{2}[\alpha_0(\mi\xi)+\alpha_0(-\mi\xi)]$ \cite{Buhmann04}.
As seen from Eq.~(\ref{eq5}), the line widths affect the
imaginary-frequency polarisability most strongly at large frequencies.
The largest impact on the CP potential is hence expected at small
distances. In Table~\ref{Tab8}, we compare the $C_3$ coefficients
including the finite line widths with those one would obtain for zero
line widths.
\begin{table*}[!t!]
\begin{center}
\begin{tabular}{|c||c|c|c|c|c|c|}
\hline
\quad\,Molecules $\rightarrow$\quad\,
&\multicolumn{3}{c|}{\quad\,$\mathrm{C}_{60}$\quad\,}
&\multicolumn{3}{c|}{\quad\,$\mathrm{C}_{70}$\quad\,}\\
\hline
\quad\,Material $\downarrow$\quad\,
&\,$C_3$\,
&\,$C_{3,\Gamma_k\to 0}$\,
&\,$C_{3,\mathrm{LRT}}$\,
&\,$C_3$\,
&\,$C_{3,\Gamma_k\to 0}$\,
&\,$C_{3,\mathrm{LRT}}$\,\\
\hline
\hline
\quad\,Perfect conductor\quad\,
&\,$2.36\times 10^{-47}$\,
&\,$2.34\times 10^{-47}$\,
&\,$2.15\times 10^{-47}$\,
&\,$2.96\times 10^{-47}$\,
&\,$2.93\times 10^{-47}$\,
&\,$2.68\times 10^{-47}$\,\\
\quad\,Au\quad\,
&\,$1.01\times 10^{-47}$\,
&\,$1.00\times 10^{-47}$\,
&\,$9.28\times 10^{-48}$\,
&\,$1.27\times 10^{-47}$\,
&\,$1.25\times 10^{-47}$\,
&\,$1.15\times 10^{-47}$\,\\
\quad\,$\mathrm{Si}_3\mathrm{N}_4$\quad\,
&\,$8.45\times 10^{-48}$\,
&\,$8.36\times 10^{-48}$\,
&\,$7.69\times 10^{-48}$\,
&\,$1.06\times 10^{-47}$\,
&\,$1.05\times 10^{-47}$\,
&\,$9.55\times 10^{-48}$\,\\
\quad\,$\mathrm{SiN}_x$\quad\,
&\,$6.26\times 10^{-48}$\,
&\,$6.21\times 10^{-48}$\,
&\,$5.75\times 10^{-48}$\,
&\,$7.86\times 10^{-48}$\,
&\,$7.76\times 10^{-48}$\,
&\,$7.11\times 10^{-48}$\,\\
\hline
\end{tabular}
\end{center}
\caption{
\label{Tab9}
Impact of absorption on the $C_3$ coefficients
[$\mathrm{J}\mathrm{m}^3$] for the CP potential of fullerene.}
\end{table*}
We note that the line widths have practically no influence on the
nonretarded potential, with differences of about $1\%$. This tiny
change is quadratic in the line widths and unobservable in an
experiment.

In an alternative approach based on linear-response theory
\cite{Henkel02}, the atomic properties enter via the
fluctuation--dissipation theorem. The resulting CP potential depends
on the atomic polarisability in its unsymmetrised form,
\begin{equation}
\label{eq26b}
C_{3,\mathrm{LRT}}=\frac{\hbar}{16\pi^2\varepsilon_0}
 \int_0^\infty\dif\xi\,\alpha_0(\mi\xi)\,
 \frac{\varepsilon(\mi\xi)-1}{\varepsilon(\mi\xi)+1}\;.
\end{equation}
The atomic transition frequencies and line widths are not considered
explicitly, but only appear at the end of the calculation when
specifying the polarisability. In particular, the line widths now
affect the potential already to linear order. As seen from
Table~\ref{Tab9}, this leads to a prediction of a reduction of the
$C_3$ coefficient by about $10\%$ due to absorption. This effect could
be visible in sufficiently accurate $C_3$ experiments, making it
possible to distinguish between the macroscopic, linear response model
for absorption and our microscopic model (which predicts that the
effect of absorption on ground-state potentials is negligible). This
possibility makes fullerenes most attractive for CP-potential studies.
In contrast, atomic systems are unable to resolve the difference
between the effects of symmetrised vs unsymmetrised polarisabilities
(or the neglect of line widths altogether).

Finally, let us discuss the impact of the infrared resonances on the
$\mathrm{C}_{60}$ potential. As seen from Table~\ref{Tab6}, their
dipole moments are much smaller than those of the optical
transitions (Table~\ref{Tab3}). On the other hand, their transition
wavelengths being much longer than those of the optical transitions,
the nonretarded limit applies over a larger range of distances. In
addition, the thermal photon numbers can take large values even at
room temperature, so that resonant potentials~(\ref{eq7}) come into
play.

At distances up to about $10\,\mu\mathrm{m}$, the CP potential due to
phonon resonances is strongly nonretarded. As shown in
Ref.~\cite{Ellingsen10}, the potential in this regime is well
approximated by
\begin{equation}
\label{eq26}
U_\mathrm{Phonon}(z)=-\frac{C_{3,\mathrm{Phonon}}}{z^3}
\end{equation}
with
\begin{equation}
\label{eq27}
C_{3,\mathrm{Phonon}}=\frac{1}{48\pi\varepsilon_0}
 \sum_{n,k}p_n(T_\mathrm{m})|\vec{d}_{nk}|^2
\end{equation}
where the sum only runs over phonon transitions. This result holds
regardless of the environment temperature for all materials of
sufficiently large permittivity. The corresponding $C_3$
coefficient depends on the internal temperature of the molecule, it
ranges from
$C_{3,\mathrm{Phonon}}\!=\!3.4\times 10^{-51}\mathrm{Jm}^3$ at zero
temperature to
$C_{3,\mathrm{Phonon}}\!=\!2.6\times 10^{-51}\mathrm{Jm}^3$ at
$T_\mathrm{m}\!=\!300\,\mathrm{K}$. A comparison with the
$C_3$ coefficients listed in Table~\ref{Tab8} reveals that the
potential contribution from infrared phonon transitions is smaller
than the discussed potential from optical transitions by more than
two orders of magnitude.

At larger distances, corrections due to imperfect reflectivity
manifest themselves \cite{Ellingsen11}. However, they do not affect
the order of magnitude of the phonon CP potential, which remains
insignificant. As we have numerically verified, the phonon
contributions only become relevant at very large distances, well
beyond $100\,\mu\mathrm{m}$.


\section{Summary}
\label{Sec4}

We have determined the Casimir--Polder interaction of
$\mathrm{C}_{60}$ and $\mathrm{C}_{70}$ with plane surfaces of Au and
two different nitride species, as commonly used in molecular
interference experiments. The numerically
calculated potentials is well approximated by the thermal, retarded
and nonretarded asymptotes for large, intermediate and small
distances, respectively. We have found that the potential is entirely
due to optical transitions and hence independent of the internal
temperature of the molecules. The environment temperature affects the
potential for distances larger than $2\,\mu\mathrm{m}$ at room
temperature. Comparing the potentials for different silicon nitride
species, we have found differences of up to $30\%$ in the nonretarded
regime, which is most relevant for diffraction experiments.

According to our microscopic theory of the molecule--field
interaction, the finite line widths of the molecules have practically
no influence on the Casimir--Polder potential. A macroscopic
linear-response approach, on the other hand, predicts that they
decrease the nonretarded potential by about $10\%$. The relatively
large line widths of the fullerene transitions thus make them an ideal
system to study the impact of molecular absorption on dispersion
interactions.


\subsection*{Acknowledgments}
We thank H.~Ulbricht and S.~Nimmrichter for discussions. Support from
the European Science Foundation (ESF) within the activity `New Trends
and Applications of the Casimir Effect' is gratefully acknowledged.
A.~J. and K.~H. acknowledge support by the MIME project within the ESF
Eurocore EuroQUASAR program.

\appendix
\section{Coefficient for the retarded Casimir--Polder potential}
\label{AppA}

The integral of the coefficient $C_4$, given in Eq.~\eqref{eq23} can
be solved explicitly with the result
\begin{align}
C_4=&\frac{3\hbar c\alpha_0}{64\pi^2\varepsilon_0}
\biggl\{\frac{10-3\sqe-4\ep-3\ep^{3/2}+6\ep^2}{3(\ep-1)}\notag \\
&-\frac{\ep^2}{\sqrt{\ep+1}}\biggl[\log\frac{\sqrt{\ep+1}-1}{\sqrt{
\ep+1}+1}+2\log\bigl(\sqe+\sqrt{\ep+1}\bigr)\biggr]\notag \\
&-\frac{2\ep^3-4\ep^2+3\ep+1}{(\ep-1)^{3/2}}
\log\bigl(\sqe+\sqrt{\ep-1}\bigr)\biggr\}
\end{align}
For large and small values of $\ep$, $C_4$ behaves like
\begin{eqnarray}
C_4&\sim& \frac{3\hbar c\alpha_0}{64\pi^2\varepsilon_0}\,
 \biggl(2-\frac{5}{2\sqe}+\frac{44}{15\ep}+...\biggr), ~~\ep\gg 1,\\
C_4&\sim& \frac{3\hbar c\alpha_0}{64\pi^2\varepsilon_0}\,
 \biggl(\frac{23}{30}\,\chi -
 \frac{169}{420}\,\chi^2+...\biggr), ~~\chi\ll 1
\end{eqnarray}
with $\chi=\ep-1$.


\end{document}